\documentclass[Afour,sageh,times]{sagej}

\usepackage{moreverb,url}

\usepackage{amsmath}
\usepackage{bm}
\usepackage{graphicx}
           
\usepackage[colorlinks,bookmarksopen,bookmarksnumbered,citecolor=red,urlcolor=red]{hyperref}

\newcommand\BibTeX{{\rmfamily B\kern-.05em \textsc{i\kern-.025em b}\kern-.08em
T\kern-.1667em\lower.7ex\hbox{E}\kern-.125emX}}

\def\volumeyear{2016}

\begin{document}

\runninghead{Iwasawa et al.}

\title{Implementation and Performance of Barnes-Hut N-body algorithm  on
  Extreme-scale Heterogeneous Many-core Architectures}

\author{Masaki Iwasawa\affilnum{1}, Daisuke Namekata\affilnum{1}
, Ryo Sakamoto\affilnum{2}, Takashi Nakamura\affilnum{2}
, Yasuyuki Kimura\affilnum{3}, Keigo Nitadori\affilnum{1}
, Long Wang\affilnum{4}, Miyuki Tsubouchi\affilnum{1}
, Jun Makino\affilnum{5}, Zhao Liu\affilnum{6}\affilnum{7}
, Haohuan Fu\affilnum{8}, Guangwen Yang\affilnum{9}}

\affiliation{
  \affilnum{1}RIKEN Center for Computational Science \\
  \affilnum{2}PEZY Computing K. K. \\
  \affilnum{3}ExaScaler Inc. \\
  \affilnum{4}Universit\"at Bonn, Argelander-Institut f\"ur Astronomie \\
  \affilnum{5}Kobe University \\
  \affilnum{6}Department of Computer Science and Technology, Tsinghua University \\
  \affilnum{7}National Supercomputing Center in Wuxi \\
  \affilnum{8}Department of Earth System Science, Tsinghua University \\
  \affilnum{9}Department of Computer Science and Technology, Tsinghua University
}


\corrauth{Masaki Iwasawa, RIKEN Center for Computational Science,
  7-1-26 Minatojima-minami-machi, Chuo-ku, Kobe, Hyogo,  650-0047, Japan}

\email{masaki.iwasawa@riken.jp}

\begin{abstract}
  In this paper, we report the implementation and measured performance of
  our extreme-scale global simulation code  on
  Sunway TaihuLight and two PEZY-SC2 systems: Shoubu System B and
  Gyoukou.  The numerical algorithm is the parallel Barnes-Hut tree
  algorithm, which has been used in many large-scale astrophysical
  particle-based simulations. Our implementation is based on our FDPS
  framework.  However, the extremely large numbers of cores of the
  systems used (10M on TaihuLight and 16M on Gyoukou) and their
  relatively poor memory and network bandwidth pose new challenges. We
  describe the new algorithms introduced to achieve high efficiency on
  machines with low memory bandwidth.  The measured performance is
  47.9, 10.6 PF, and 1.01PF on TaihuLight, Gyoukou and Shoubu System B
  (efficiency 40\%, 23.5\% and 35.5\%). The current code is developed
  for  the simulation of  planetary rings, but most of the new
  algorithms are useful for other simulations, and are now available
  in the FDPS framework.
\end{abstract}


\maketitle

\section{Introduction}
\label{sect:overview}

The architecture of HPC platforms has shown significant changes in the
last three decades, from vector-parallel architecture to
distributed-memory scalar processors, and then many-cores with SIMD
units, accelerators, and most recently heterogeneous many-cores.

The ideas behind the accelerators and heterogeneous many-cores are
very similar. In both cases, a number of relatively simple and thus
energy- and area-efficient processors are combined with relatively
small number of complex, high-performance processors. The difference
is that in the case of the accelerator systems, the complex,
high-performance cores are usually commodity processors (the host CPU)
in one die, and accelerator cores are in a separate die, and they are
connected by general-purpose link such as the PCI express. The host
CPU is usually of x86 architecture.  GPGPUs used with x86 processors
are the currently the most widely used accelerator systems. This
architecture has many advantages, but the most important one is that
all of the hardware and software for the host CPU is already
there. The developer of the accelerator hardware can concentrate on
the accelerator hardware itself and software to make use of it.

The disadvantage of the accelerator system is the existence of the
communication link between the host and accelerators. Usually, both
the bandwidth and latency of the communication between the host and
accelerators is limited by that of the standard PCIe specification.
The theoretical peak throughput of the PCIe 3.0 standard (with 16
lanes) is only 16GB/s. On the other hand, some of latest GPGPUs have
multiple channels of the HBM memory, with the total bandwidth
approaching to 1TB/s.  If we divide the calculations of an application
into that on the host and that on accelerators, communication between
them would become necessary, and in many cases that would limit the
performance. Thus, either we have to port all of application to the
accelerator side, or we have to live with relatively low performance.
In addition, in many cases size of the on-board memories of the
accelerator boards is small, less than 16GB, and thus either the
problem size is limited or we need to store the data in the main
memory of the host CPU.

In principle, heterogeneous many-cores can solve these limitations of
accelerator systems, since what correspond to the host CPUs of
accelerator systems are now integrated to the same LSI chip as the
accelerators, and they share the same physical memory.
Examples of such architecture include  Sunway SW26010 and PEZY-SC2.
The former is used in the TaihuLight system, which was ranked \#1 in
the Top 500 list four times in years 2016 and 2017. The latter was
used in the Gyoukou system, which was ranked \# 4 in the Top 500 list
of November 2017. We can see that though the heterogeneous many-core
architecture has clear advantages, they are yet to be widely used.

Currently, to port applications to these systems and achieve good
performance requires quite a lot of efforts. One reason is that
``automatic'' parallelization through, for example, OpenACC results in
rather poor performances in the case of TaihuLight, and currently only
a subset of OpenCL is available on PEZY-SC2. In the case of PEZY-SC2,
right now the host CPU on chip is disabled and thus the current system
an accelerator system with a Xeon-D CPU. However, since the memory on
PEZY-SC2 is larger than that on Xeon-D, all data can (and should) be
on the side of PEZY-SC2.

Another reason is that the performance ratio between the complex cores
and simple cores tend to be very large. In the case of TaihuLight, one
complex core and one simple core have the same peak performance, but
there is only one complex core for every 64 simple cores. Thus, even
though it is not easy to write codes for simple cores, almost all
codes should be moved to the simple core side to obtain decent
performance.

One way to make the application development on these machines easier
is to provide DSLs or frameworks, in which the users express the
problem to be solved or numerical method to be used in high-level,
machine-independent way. The machine-specific part of DSL runtime
library must be optimized to each architecture, but since one DSL can
be used to implement many applications, if such a DSL is possible a
lot of works by many researchers can be used for more productive
researches.

We have been developing FDPS, Framework for Developing particle
simulators \citep{Iwasawaetal2016}.  The basic idea of FDPS is to
provide a set of library functions necessary for high-performance,
highly scalable particle simulation codes. FDPS receives from user
applications the definition of particle data (originally in C++ class)
and a function to evaluate particle-particle interactions as the
source code. FDPS itself is written as a template library which is
compiled with these user-defined class and interaction function. Thus,
a user's application can call functions defined in FDPS to process
particles they defined and to calculate the interactions they
defined. We have extended the API so that FDPS can accept the particle
data class (or struct) and the interaction function written in both
Fortran \citep{2018PASJ...70...70N} and C. The pure-C language
interface makes it possible for programs written in any language with
reasonable FFI to C language to use FDPS functions.

FDPS relies on parallel Barnes-Hut tree algorithm with domain
decomposition by multisection algorithm and local essential tree
method \citep{Makino2004} for interaction calculation. Currently, FDPS
supports usual multicore architectures and also accelerator
architectures. In the case of the support of accelerator
architectures, the multiwalk algorithm \citep{Hamadaetal2009} is
used. This means that everything other than the interaction
calculation using the interaction list is done on the host side, and
the user-supplied interaction function need to take care of the data
transfer between the host and accelerators.

Thus, currently FDPS does not support heterogeneous many-cores very
well. One could  use the  accelerator support, but the performance gain
would be rather limited because of the reasons described above.
In order to improve the efficiency, it is necessary to move operations
other than the interaction calculation such as the tree construction
and the construction of the interaction list to the simple cores. 

In this paper, we report the result of porting relatively simple
N-body simulation code for planetary ring systems, developed based on
our FDPS framework, to two heterogeneous many-core processors: Sunway
SW26010 and PEZY-SC2. The code is not yet the complete port of FDPS to
these processors, but more like a production code based on FDPS for a
specific problem. The reason why full FDPS port is not yet done is
simply that we made this porting partly to evaluate the architecture
and partly to try large-scale calculations which were not practical on
other architectures.

In the rest of this paper, we
first give an overview of the current state of the arts for the
large-scale simulations of planetary rings in section \ref{sect:rings}, and then 
short description of the architecture of the two systems
in section \ref{sect:hardware}.  In section \ref{sect:innovation}, we
describe in detail the new algorithms we developed to achieve high
performance on extreme-scale heterogeneous many-core architectures.
In section \ref{sect:performance} we describe how the performance was
measured and achieved performance.  Section \ref{sect:summary} is for
discussion and summary.

\section{Simulation of planetary rings}
\label{sect:rings}

Saturn's ring was first observed by Galileo Galilei in 1610. For more
than three centuries, it had been the only known ring system within
our solar system. In 1977, rings of Uranus were found through
occultation observations from an aircraft, and then in 1979 rings of
Jupiter by Voyager 1 and in 1989 those of Neptune by Voyager 2.  Very
recently, it turned out that some of minor planets also have
rings. The first distinctive example is 10199 Chariklo, whose orbit is
between those of Saturn and Uranus (and thus one of Centaurs). There
are probably more Centaurs with rings.

Thus, quite recently, a wide variety of ring systems have been found.
How these rings were formed and have evolved is an important question
in planetary science, and large-scale, global simulation, if possible,
would help greatly to advance our understanding.

Planetary rings are usually at the radii around the Roche limit. Thus,
mutual gravity between particles does not easily lead to the formation
of new satellites, but is important enough to form spiral waves
(``wakes'') in very small scales, which increase the effective
viscosity and should enhance the radial transport of the angular
momentum. On the other hand, the actual ring system seems to consist
of very large number of narrow rings, separated with distinct gaps. It
is believed that these gaps are maintained by high-order resonances
with small embedded satellites (so-called moonlets), but whether or
not such gaps can be formed and maintained by resonances has not been
fully understood.

Up to now, most of simulations of ring structures have been local
ones, in which a small patch was cut out from the ring and simulated
under the assumption of the local Hill approximation and periodic
boundary condition \citep{WisdomTremaine1988}. Rein and Latter
\citep{ReinLatter2013} performed ``Large-scale'' simulation of viscous
overstability in Saturn's rings, using up to 204,178 particles and up
to 10,000 orbits using this local approach.  Because very long
simulations are necessary, the number of particles has been
small. They used {\tt REBOUND} \citep{ReinLiu2012}, an MPI-parallel
$N$-body simulation code.

Michikoshi and Kokubo \citep{MichikoshiKokubo2017} performed global
simulations of rings with the largest number of particles reported so
far. They used 300M particles to model two narrow rings of
Chariklo. They have developed their parallel code using the framework
we developed, FDPS \citep{Iwasawaetal2016, 2018PASJ...70...70N}.

Almost all previous studies of planetary rings adopted so-called
``local'' approximation, in which only a small patch of a ring is
simulated assuming periodic boundary condition in both radial and
azimuthal directions.

Michikoshi and Kokubo \citep{MichikoshiKokubo2017} performed global
simulations of rings with 300M particles, using FDPS
\citep{Iwasawaetal2016, 2018PASJ...70...70N}. They so far followed the system only for 10
orbital periods.

The total calculation cost is roughly proportional to number of
particles multiplied by the number of orbital periods followed, since
the calculation cost per timestep is $O(N \log N)$ when Barnes and Hut
tree algorithm is used and the number of timestep required for ring
simulations is essentially independent of the number of
particles. Thus, we can conclude that the size of state-of-the-art
simulations of planetary rings is around $10^9$ particle-orbits, or
around $10^{12}$ particle-steps.

We should note that even though the simulations so far done in this
field is relatively small, that does not mean there is no need or
possibilities for larger scale simulations. If we want to model the
global structures of rings, we cannot rely on local treatment. For
example, the effect of resonances with small satellites can only be
studied using global simulations. On the other hand, the number of
particles one need for global simulations, even for a very narrow
radial range, is very large. For example, consider the A ring of
Saturn with the radius of around $1.3\times 10^5\rm km$. The typical
radius of ring particles is 6~m \citep{ZEBKER1985531}, and the optical
depth of the ring is around unity. Thus, we need $10^4$ particles per
square km or around $10^{12}$ particles for the radial range of 100
km. With this radial range, we can model many of fine features
observed by Cassini directly.

If we could use particles with larger size, we could reduce the number
of particles required significantly. However, that would change the
viscous diffusion timescale of the ring, and thus what would be
observed. It is necessary to perform
simulations with particles of real physical radius, which would require
at least $10^{16}$ and ideally $10^{19}$ particle steps.

In other fields of astrophysics, very large simulations have been
performed. For example, Ishiyama \citep{Ishiyama2014} used $4096^3$
particles to follow the formation and growth of dark matter halos of
smallest scales. This simulation corresponds to $10^{16}$ particle
steps. Part of this calculation was performed on K computer. The
performance of K computer is $4.0\times 10^{10}$ particle steps per
second on the entire K computer, or 60,000 particle step per second
per core for a processor core with the theoretical peak performance of
16 Gflops \citep{Ishiyamaetal2012}. The efficiency they achieved is
55\% of the theoretical peak.

The algorithms used in large-scale $N$-body simulations are rather
similar, and that means they are well studied and close to optimal.
All of them use domain decomposition and Barnes and Hut tree
algorithm. For domain decomposition, several variations have been
used, such as Orthogonal Recursive Bisection \citep{Salmon1990},
Hashed Oct Tree \citep{WarrenSalmon1992}, Multisection
\citep{Makino2004}.

Efficient implementations on large-scale GPGPU clusters exist
\citep{Hamadaetal2009, PortegiesZwartetal2014, Bedorfetal2014}.
\citet{Bedorfetal2014} performed the simulation of Milky Way Galaxy
using $2.42 \times 10^{11}$ particles. The achieved performance is
24.77 PF on ORNL Titan, and one timestep took 5.5 seconds. Thus they
have achieved the performance of $4.4 \times 10^{10}$ particle steps
per seconds. The theoretical peak performance of Titan is 73.2 PF in
single precision. Thus, the achieved efficiency is 33.8\%.


\section{Sunway TaihuLight and PEZY-SC2 systems}
\label{sect:hardware}

In this section, we briefly describe the features of the Sunway
TaihuLight system and two systems with PEZY-SC2 processors: Gyoukou
and Shoubu System B. For more details of TaihuLight system see
\citet{Fu2016}. TaihuLight consists of 40960 Sunway 26010 processors,
and Gyoukou and Shoubu System B 13312 and 512 PEZY-SC2 processors,
respectively. Unfortunately, Gyoukou was turned off by March 31, 2018,
and thus our performance measurement on Gyoukou system was based on a
preliminary version of the simulation code, and the efficiency
measured on Gyoukou is lower than that measured on Shoubu System B.


One SW26010 processor consists of four  CGs (core groups), each with one
MPE (management processing element) and 64 CPEs (computing processing
elements). Both MPE and CPE are 64-bit RISC cores. MPE has L1 cache
memories for  both instructions and data, and also L2 data cache. On
the other hand, each CPE has L1 instruction cache and 64KB of local
data memory. CPEs can communicate with the main memory through DMA.
Each CPE can initiate multiple asynchronous DMA operations. 

Each core group is connected to 8GB DDR3 DRAM memory with the theoretical
peak transfer rate of 34GB/s. The processor runs at the clock
frequency of 1.45GHz, and each core (both MPE and CPE) can perform
four double precision FMA operations. Thus, the theoretical peak
performance of one processor is 3016 Gflops and that of one CG is 754
Gflops. Thus, even when we use the nominal number for DRAM bandwidth,
the B/F ratio is only 0.045. This is less than 1/10 of the number for
K computer.

Compared to that of K computer, the network is also  weak,
with the total bandwidth of around 10GB/s per node. This is about the
same as the performance of a single link of 6D torus network of K
computer. Since the SW processor is around 25 times faster than the
SPARC64 processor of K computer, the relative network bandwidth is
different by  more than two orders of magnitudes.

One PEZY-SC2 processor chip consists of  2048 processors (64
of them are disabled and the available number of processors is
1984). Each processor can perform 1, 2 and 4 multiply-and-add
operation for FP64, FP32, and FP16 data. For FP32 and FP16, 2- and
4-way SIMD operations are performed.  With the clock speed of 700MHz,
the theoretical peak speed is 2.8, 5.6 and 11.1TF, for FP64, FP32 and
FP16, respectively. At present, each SC2 processor chip have 4 channels of
DDR4 memory, for the peak throughput of 76.8GB/s. Thus B/F is 0.027.

They have three levels of shared cache, but without
coherency. Instead, they have explicit cache flush instructions to each
levels. Two processors share L1D, and 16 processors L2D, and all
processors LLC. Each processor runs either four or eight threads
simultaneously. Thus, it is relatively easy to hide the  latency of
the arithmetic units and L1D.

In the original design, each SC2 processor chip had six MIPS64 cores,
which were supposed to run the operating system and main body of the
application programs. Unfortunately, currently they are disabled, and
operating system and application programs run on the frontend  Xeon~D-1571
processor. Each Xeon~D  hosts eight SC2 processors. Thus, the
performance ratio between Xeon~D and SC2 is close to 100.
Moreover, these eight SC2 are connected to Xeon~D through single PCIe
Gen3 16-lane channel. Thus, the peak data transfer speed between one
SC2 and Xeon~D is 2 GB/s, for the peak speed of 2.8TF.

%
In summary, TaihuLight and two systems based on PEZY-SC2 processors  share the following
characteristics:

\begin{enumerate}

\item Very large performance ratio between ``general-purpose''
  and ``computing'' cores, close to 1:100.
\item  Very small memory  B/F numbers, around 0.03.
\item  Even smaller network B/F numbers, 0.006 or 0.001.
\item Very large number of MPI processes, 160k  or 10k.
\item Very large number of ``computing'' cores per MPI process, 64 or 1984.  
\end{enumerate}

Just one of these characteristics makes it very difficult to achieve
reasonable performance for particle-based simulations using previously
known parallelization algorithms. In the next section, we describe the
new algorithms we implemented to achieve good performance on these systems.

\section{New algorithms for extreme-scale simulations}
\label{sect:innovation}

  \subsection{Overview of new algorithms}

In this section, we describe the new algorithms we made in order to
utilize TaihuLight and PEZY-SC2 based systems for the simulations of self-gravitating
planetary rings. The following is the list of new algorithms.

\begin{enumerate}
\item The reuse of the interaction list over multiple timesteps.
\item Elimination of  the global all-to-all communication.
\item ``Semi-dynamic'' load balance between computing cores
\item Optimizations specific to the ring geometry.
\end{enumerate}  

For TaihuLight and PEZY-SC2 based systems, we have modified our FDPS
framework in architecture-specific way so that we implement the
algorithms and run the code under the limited available time and
software environment. However, many of these algorithms are ported
back to the original FDPS so that anybody who uses FDPS can take
advantage of these new algorithms.

In the rest of this section, we briefly describe these new innovations.

\subsection{Reuse of the interaction list}
\label{subsec:list}

The following gives the usual steps for highly parallel code for
self-gravitating particle system:

\begin{enumerate}

  \item Perform domain decomposition.
  \item Exchange particles so that particles belong to appropriate domains.
  \item Perform interaction calculation using fast algorithm such as
    Barnes-Hut tree.
  \item Integrate the orbits of particles.
  \item Go back to step 1.

\end{enumerate}

In the case of approaches with local essential tree, step (3) consists
of the following substeps:

\begin{description}

\item{(3a)} Construct the ``local'' tree structure from particles in
  the domain.
\item{(3b)} Collect the information necessary for the calculation of
  interaction from other processes (so called local essential tree).
\item{(3c)} Construct the ``global'' tree from the collected information.
\item{(3d)} For small groups of particles, traverse the tree and
  calculate the interaction. Repeat this for all groups.
\end{description}
In the original algorithm \citep{BarnesHut1986}, the traversal of the
tree is done for each particle, and force calculation is done during
the traversal. However, on almost all modern implementation, following
the idea of Barnes \citep{Barnes1990}, tree traversal are done for
groups of neighboring particles, which are constructed using the tree
structure itself. During the traversal for a group, the list of
particles and tree nodes which exert the force on this group of
particles is constructed, and actual force calculation is done through
the double loop over particles in the group and those in the
list. This structure makes it possible to use vector pipelines, scalar
SIMD units, and even special-purpose computers \citep{Makino1991c}
with high efficiency. For GPGPUs, the extension of this algorithm, in
which multiple lists are constructed and then sent to GPGPU, is used
\citep{Hamadaetal2009}.

This approach does not work well on TaihuLight or PEZY-SC2 based systems, because of
the low performance of general-purpose core and limited memory bandwidth. 
The
performance we can achieve with either approach for ring simulation on
these machines is less than 1\%.
Thus, it is necessary to reduce the cost of tree construction and tree
traversal, and we achieved this by using the same interaction lists
over multiple timesteps. We call this method the persistent
interaction list method.

The idea behind this method is essentially the same as that for
the  neighbor-list method used in many simulation codes for particles
with short-range  interactions. 

By using this persistent interaction list, we can reduce the
calculation cost of the part other than the interaction calculation
drastically. While we are using the same interaction lists, we skip the
domain decomposition, exchange of particles, construction of the local
tree. We still need to update the physical quantities of the nodes of
the tree, since particles move at each timestep.
We first update the information of the nodes of
the local tree. Then,
using the list of nodes for the local essential tree, the
communication between the nodes is performed. Finally, the physical
quantities of the global tree are updated, and the force calculation is
performed using this updated global tree and the persistent
interaction list.

The most time-consuming part of the tree construction is the
sorting. In the case of TaihuLight, we implemented the parallel sample
sort \citep{cmsort} on CPEs. In the case of Gyoukou, the sorting was
performed on Xeon~D host processor. In the case of Shoubu System B, it
was performed on the side of PEZY-SC2 processors. Also, some other
operations are moved from Xeon-D to PEZY-SC2. Thus, the overall
performance is significantly better for Shoubu System B. As stated
earlier, Gyoukou was turned off on March 31, 2018, and we could not
measure the performance of our improved code on systems with more than
512 PEZY-SC2 processors.

We have ported all operations in timesteps in which the interaction
list is used (list-reusing steps), except for MPI functions for
communication, to CPEs (TaihuLight) or SC2 processors (PEZY-SC2 based
systems).  For the timestep in which the interaction list is
constructed (list-constructing step), some of operations are still
done on Xeon~D in the case of PEZY-SC2 based systems.

\subsection{Tree and Domain structures on Cylindrical Coordinate}
\label{subsec:cylcoord}

\begin{figure}
  \centering
  \includegraphics[width=8cm,clip]{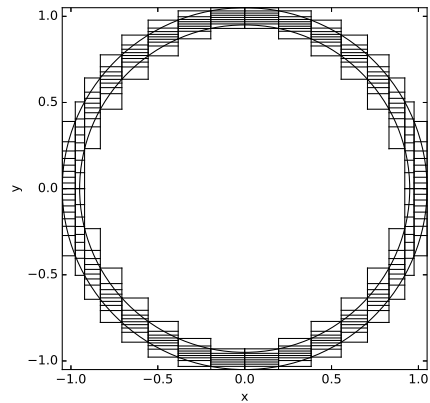}
  \caption{Schematic figure of domain decomposition by the multisection
    method in $x$-$y$ coordinate. Domains are divided by $16 \times 16$.}
  \label{fig:domain_cart}
\end{figure}

We want to model a relatively narrow ring, and this means the usual
domain decomposition in Cartesian coordinates can cause serious
problems. Figure~\ref{fig:domain_cart} illustrates the problem. We can
see the domains near the $y$ axis are very elongated. This irregular
shape of domains results in the increase of communication between
processes, and thus serious degradation in the efficiency.

\begin{figure}
  \centering
    \includegraphics[width=8cm,clip]{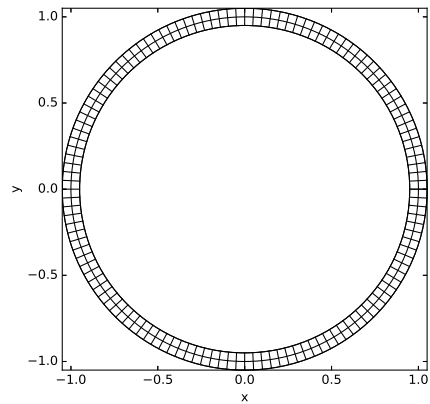}
  \caption{Schematic figure of domain decomposition by the multisection
    method in cylindrical coordinate. Domains are divided by $2 \times 128$.}
  \label{fig:domain_cyl}
\end{figure}

We can avoid this problem, if we apply the domain decomposition 
in the  cylindrical coordinates (figure  \ref{fig:domain_cyl}).
Note that we can also use the  cylindrical coordinates for the
construction of the tree.  Since
the ring is narrow, the local distance $s$ in the Cartesian coordinates
$(x, y, z)$ can be approximated by that in cylindrical coordinates
($r$, $\phi$, $z$).
\begin{equation}
  \label{eq:metric}
  ds^2 = dx^2 + dy^2 + dz^2 \sim d\phi ^2 + dr^2 + dz^2,
\end{equation}
when $r \sim 1$. Thus,
we can use the cylindrical coordinate
for domain decomposition and tree
construction and even for the tree traversal,  without any modification of the algorithm or program
itself. The actual interaction calculation is faster in Cartesian
coordinates and thus Cartesian coordinates is used.

\subsection{Coordinate rotation}
\label{subsec:exptcl}

The simulation of ring with very large number of processes poses
new challenges. As we increase the number of processes, the size of
the domains becomes smaller. On the other hand, the timestep does not
become much smaller even when we increase the total number of
particles, since the random velocities of ring particles become
smaller when we increase the number of particles. Thus, the distance
that particles move can be comparable or even larger than the domain
size, resulting in the increase in the amount of  communication.

We can ``solve'' this problem by the rotation of  the coordinates and domain
structure, so that particles do not move much. If we rotate the
coordinates at the speed of Kepler rotation at the center of the ring,
particles at the center of the ring do not move much. Particles at
other radial positions still move, but the speed becomes much smaller
than that of the Kepler rotation. Thus, communication due to Kepler
rotation can be almost eliminated.

\subsection{Elimination of all-to-all communication}
\label{subsec:exlet}



In FDPS, the exchange of LET (local essential tree) data is done
though a single call to the {\tt MPI\_Alltoallv} function.  This
implementation works fine even for full-node runs on K computer, but
becomes problematic on systems with relatively weak network like
TaihuLight and PEZY-SC2 based systems. We can eliminate this all-to-all communication, by
constructing the ``tree of domains'' locally and let only higher-level
information be sent to distant processes.

In the current implementation specialized to narrow rings, we
implemented a very simple two-level tree, in which the second-level
tree nodes have all processes in the radial direction. For example, if we
have a process grid of (1000, 10), where 1000 in angular and 10 in
radial direction, 10 domains in the radial direction are combined to
one tree node, resulting in 1000 second-level nodes. Only these 1000
nodes exchange their center-of-mass information. All LET information
other than these center-of-mass data of second-level nodes are sent
either to other second-level nodes (and then broadcast to
lower-level nodes) or sent directly to lower-level nodes.

In this implementation, there is still one global communication in the
angular direction, but we can use {\tt MPI\_Allgather} since only the
top-level data are sent. Thus the reduction in the communication was
quite significant.

\subsection{Load Balance among computing cores}
\label{subsec:force}

In our current implementation, interaction lists are created at the
list-construction step, and are reused for several steps. The total
number of lists in one MPI process is around $10^5$, and we need to
use 64 or 1984 computing cores efficiently for them.
If we just assign a fixed number of lists to cores, random variation
of the list length can result in large load imbalance. Therefore, some
load balance strategy is necessary. We applied the following simple algorithm.

\begin{enumerate}
  
\item Sort the interaction lists by their length.

\item Assign the longest 64 lists on 64 CPEs (in case of TaihuLight).

\item For each remaining list, assign it to the the CPE with the
  shortest total calculation cost.
  
\end{enumerate}

Since the calculation time of cores is quite predictable, this
algorithm works very well.

In the case of PEZY-SC2 based systems, we further improved the load
balance by using multiple cores which share the cache for one
interaction list.

\subsection{Interaction Kernel}

In the case of TaihuLight, we found the compiler-generated code for
the interaction kernel, even when SIMD operations are used, does not
give very good performance. We rewrite the interaction kernel fully in
the assembly language, with hand-unroll and careful manual scheduling. As
a result, we achieved more than 50\% of the theoretical peak
performance for the kernel.

We have applied similar optimization also on PEZY-SC2 based
systems. In addition, on PEZY-SC2 based systems we used
single-precision calculation for the interaction kernel. In order to
avoid the large roundoff at the first subtraction of the position
vectors, both positions and velocities are shifted with the new origin
at the position of one of the particles which share the interaction
list.  After this shifting, positions and velocities are converted to
single precision, and actual interaction calculation is done using
single-precision SIMD operations.

\section{Measured performance}
\label{sect:performance}
\subsection{How the performance is measured}

To measure the performance, we measure the time for 64 timesteps,
including the time for  diagnostics. 
The
execution time is measured by the MPI wallclock timer, and operation
count is from the counted number of interactions
calculated. Equation~\ref{eq:interation}  gives the definition of the
particle-particle interaction. 

{
\begin{equation}
  \bm F_{ij} = \begin{cases} G \dfrac{m_i m_j}
    {r_{ij}^3} \bm r_{ij} & \left(r_{ij} > r_\text{coll} \right)
    \\
    \Biggl[  G \dfrac{m_i m_j} {r_\text{coll}^3}  + \dfrac{m_j}{m_i
        + m_j}  \times \\ \qquad  \left(      \kappa \dfrac{r_{ij} -
        r_\text{coll}}{r_{ij}}    + \eta \dfrac{\bm r_{ij} \cdot \bm
        v_{ij}}{r_{ij}^2}    \right) \Biggr] \bm r_{ij} & \left(
    r_{ij} \le r_\text{coll} \right) \end{cases}
  \label{eq:interation} 
\end{equation}
}
with
$\bm r_{ij} = \bm r_j - \bm r_i$, $\bm v_{ij} = \bm v_j - \bm v_i$,
$r_{ij} = \| \bm r_{ij} \|$

Here, ${\bm F_{ij}}$ is the acceleration of particle $i$ due to
particle $j$, ${\bm r_{ij}}$ and ${\bm v_{ij}}$ are the
relative position and velocity vectors, $G$ is the gravitational
constant (taken to be unity in this paper), $m_i$ is the mass of
particle $i$,  $r_\text{coll}$ is the distance at which
two particles collide, and $\eta$ and $\kappa$ are parameters which
determine the coefficient of restitution. We chose these parameters 
so that the coefficient of restitution in radial direction is 0.5.

We used this form to calculate all particle-particle interaction. For
particle-tree-node interaction, we used center-of-mass
approximation. Particle-particle interaction consists of 9
multiplications, 8 additions, and one square root and one division
operations. Instruction set of Sunway 26010 processor does not include
fast approximation for neither square root or reciprocal square
root. So we implemented fast initial guess and high-order convergence
iteration in software. The number of operations in this part is 7
multiplications, 5 additions and two integer operations. Therefore,
for particle-cell interactions the number of floating-point operations
is 31, and for particle-particle interactions, which include the
repulsive force during physical collisions, is 49.  The total number
of floating-point operations is obtained by counting the number of
interactions calculated and multiply them with these number of
floating-point operations per interaction. We ignore all operations
other than the interaction calculation, since as far as the number of
floating-point operations is concerned, that for interaction
calculation is more than 99\% of total operation count.

For PEZY-SC2 based systems we used the same operation count as we used
for TaihuLight, in order to make the direct comparison possible, even
though the details of the implementation of the force kernels are
different.

For the weak-scaling measurement, we have performed runs with 10M
particles per MPI process on TaihuLight and PEZY-SC2 based
systems. Initial condition is such that the ring width and ring radius
is unchanged. Table \ref{tab:initialcoditions} summarizes the initial
condition.

\begin{table}
\centering
 \caption{Initial condition for weak scaling runs}
 \label{tab:initialcoditions}
 \begin{tabular}{lc}
\hline
   Central planet & Saturn\\
   Ring inner radius & $10^5$ km\\
   Ring width        & $100$ km\\
   Number of MPI processes & 1024 -- 160,000 \\
   Number of particles per process & $10^7$  \\
   particle radius & 3.5 -- 500 m\\
\hline
\end{tabular}
\end{table}
  
\subsection{Performance Results}
\begin{figure}
  \includegraphics[width=8cm]{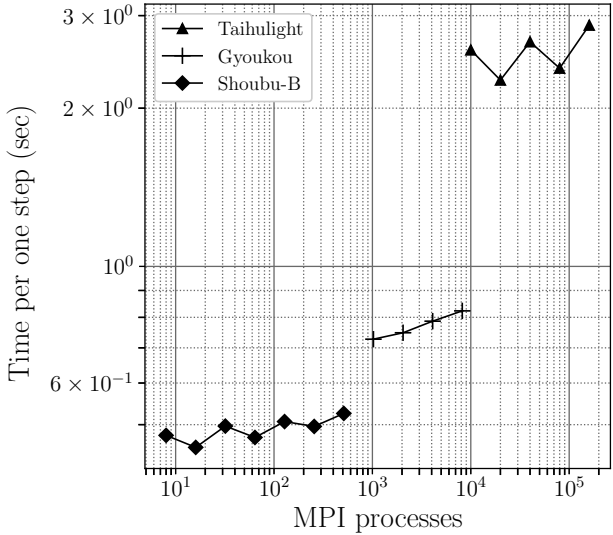}
  \caption{Time per timestep for weak-scaling test. The number of
    particles per process is 10M. Triangles, crosses and squares show the
    results on TaihuLight, Gyoukou and Shoubu System B, respectively.
  }
  \label{fig:weak}
\end{figure}

\begin{figure}
  \centering
  \includegraphics[width=8cm, clip]{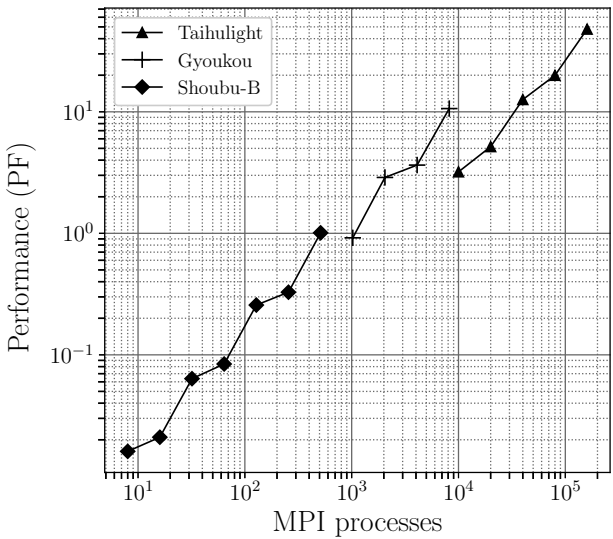}
  \caption{Performance in petaflops for weak-scaling test.  The number of
    particles per process is 10M. Triangles, crosses and squares show the
    results on TaihuLight, Gyoukou and Shoubu System B, respectively.}
  \label{fig:weakpf}
\end{figure}

\begin{figure}
  \centering
  \includegraphics[width=8cm, clip]{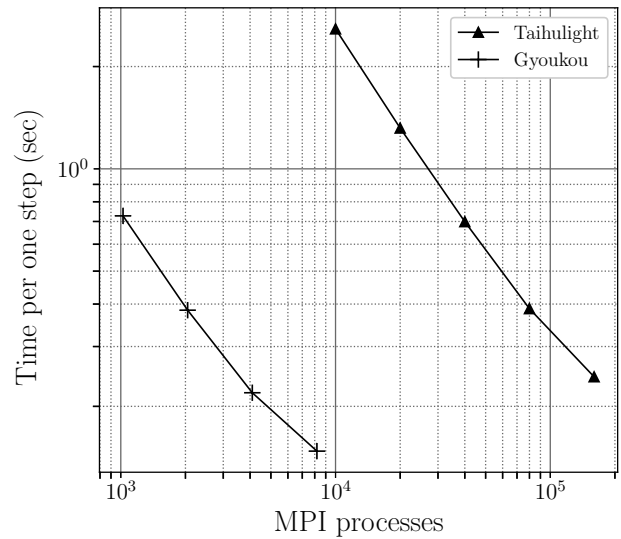}
  \caption{Time per timestep for strong-scaling test.  Triangles and crosses  show the
  results on TaihuLight and PEZY-SC2 based systems, respectively. The total number of
  particles is $10^{11}$ on TaihuLight and $10^{10}$ on PEZY-SC2 based systems.}
  \label{fig:strong}
\end{figure}

Figures \ref{fig:weak} and \ref{fig:weakpf} shows the time per one
timestep and the performance, for the weak scaling measurements. We
can see that the weak scaling performance is quite good on both of
TaihuLight and PEZY-SC2 based systems. The peak performance of single
MPI process is roughly four times faster for PEZY-SC2 based systems,
and that's the reason why they are three to four times faster in this
weak-scaling measurement. We can see that Shoubu System B is about
50\% faster than Gyoukou. As we have already discussed, this is not
due to any hardware difference but the difference in the software
used.

Figures \ref{fig:strong} shows the strong scaling result.  The total
number of particles is $10^{11}$ and $10^{10}$, for TaihuLight and
Gyoukou. We do not show the strong-scaling result for Shoubu System B,
since it is rather small system and strong-scaling result is not so
meaningful. 
We can see that speedup is almost linear.

\begin{table}
\centering
  \caption{Breakdown of calculation time for weak-scaling runs}
  \label{tab:timeweak}
  \begin{tabular}{ccccc}
    \hline
 System  & \# processes & interaction & comm. & others\\
    \hline  
&10000 &   2.07& 0.041 & 0.466\\
&20000 &  1.63& 0.040  & 0.590 \\
TaihuLight &40000  &  2.13 & 0.064 & 0.478 \\
&80000 &   1.71 & 0.053 & 0.630\\
&160000 &  2.31& 0.090& 0.476\\
\hline
&1024&  0.332 &  0.114 & 0.281\\
Gyoukou &2048&   0.392 & 0.121 & 0.235\\
&4096&  0.355 & 0.143& 0.289\\
&8192   & 0.453& 0.147&     0.222\\
\hline
        &8&  0.327 &  0.018 & 0.132\\
Shoubu B &32&   0.344 & 0.020 & 0.181\\
        &128&  0.348 & 0.027& 0.132\\
        &512& 0.360& 0.030&     0.135\\
\hline
\end{tabular}
\end{table}

\begin{table}
\centering
  \caption{Breakdown of calculation time for strong-scaling runs}
  \label{tab:timestrong}
  \begin{tabular}{ccccc}
    \hline
 System  & \# processes & interaction & comm. & others\\
    \hline  
& 10000& 2.0738 & 0.0410 & 0.4658 \\
& 20000& 1.0499 & 0.0253 & 0.2426 \\
TaihuLight & 40000& 0.5565 & 0.0298 & 0.1125 \\
& 80000& 0.2991 & 0.0233 & 0.0652 \\
& 160000& 0.1765 & 0.0322 & 0.0356 \\
\hline
& 1024& 0.3323 & 0.1140 & 0.2808 \\
Gyoukou & 2048& 0.1512 & 0.0668 & 0.1658 \\
& 4096& 0.0854 & 0.0417 & 0.0923 \\
& 8192& 0.0538 & 0.0357 & 0.0582 \\
    \hline  
\end{tabular}
\end{table}
 
Tables \ref{tab:timeweak} and \ref{tab:timestrong} show the breakdown
of the calculation time per one timestep, again for both the weak and
strong scaling runs. As expected, in the case of strong-scaling runs,
the calculation time for communication does not decrease
significantly, and eventually limits the performance.  As already
stated, our main interest is to use very large number of
particles. Therefore, for actual scientific runs, the communication
time would not become the limiting factor.

If we compare the calculation times on Gyoukou and Shoubu System B, we
can see that the times for the interaction calculation are
similar. but for both communications and ``others'', Shoubu System B is
much faster. Again, this is not due to hardware difference but due to
software difference.

The performance of run for $1.6\times 10^{12}$ particles on 160k
processes (40000 nodes) of TaihuLight is 47.9 PF, or 39.7\% of the
theoretical peak performance of the Sunway TaihuLight system.  On
PEZY-SC2 based systems, we achieved 10.6PF for $8\times 10^{9}$
particles on 8K SC2 chips, or efficiency of 23.3\% of the theoretical
peak performance. On 512-chip Shoubu System B, we achieved the speed
of 1.01 PF, or 35.5\% 

The overall efficiency we achieved on PEZY-SC2 based systems is a bit lower compared
to that on TaihuLight. This difference is not due to any fundamental
difference in the architecture but purely due to the limitation on the
available time for program development and performance measurement. As
we stated, the calculation in the list-construction step, such as
the constructions of the tree and the interaction lists are currently
done on Xeon~D, and around 40\% of the total time is consumed in this
step at the time of measurement on Gyoukou. Most of these are now done
on SC2 side, and that is why the performance of Shoubu System B is
better than that of Gyoukou.

In terms of the number of particles integrated per second, we have
achieved $5.5\times 10^{11}$ particles per second, which is more than
10 times faster than the results of previous works on K computer
\citep{Ishiyamaetal2012} or ORNL Titan \citep{Bedorfetal2014}.

\section{Discussion and summary}
\label{sect:summary}

\subsection{Performance Portability}

We have reported the measured performance of two rather different HPC
systems, Sunway TaihuLight and PEZY-SC2 based systems, for the same
large-scale simulation of self-gravitating planetary rings.  In both
cases, we have achieved fairly high efficiency, more than 30\% of the
theoretical peak. The parallel algorithm used is essentially the same
for the two systems. However, the actual codes are rather different,
simply because of the difference in the architecture and the software
development environment.

Sunway TaihuLight has a heterogeneous many-core architecture
integrated in one chip. Thus, the CPU (MPE in their terms) and
accelerators (CPE in their terms) share the same physical memory, but
CPEs lack the data cache and need to rely on DMA controller to access
the main memory efficiently.

On TaihuLight, one can use OpenACC compiler. However, in order to
achieve high performance, one is practically forced to use the Athread
call, which makes the 64 CPEs and their local memories visible to
programmers.

On the other hand, PEZY-SC2 systems, at least at present, have a
rather classical accelerator-based architecture, in which CPU (a
Xeon-D processor) and accelerators (PEZY-SC2 processors) are connected
through PCI Express interface. This means that they have separate
physical memories. Within one chip, however, processing elements of
PEZY-SC2 processor have three levels of data caches. Currently PZCL, a
dialect of OpenCL, is supported on PEZY-SC2 based systems.

Because of these differences (shared and separate memory, DMA and
cache, thread-based and OpenCL-like), the actual programs for two
machines have become quite different, even though the algorithms used
are the same and the problem to be solved is the same.

Both codes, however, are based on our framework, FDPS
\citep{Iwasawaetal2016, 2018PASJ...70...70N} and follow its
structure. The basic idea of FDPS is to separate the implementation of
parallel algorithms and description of the physical problem. FDPS
provides the former and the application programmers provides the
latter, in the form of the data type definition of particles and
functional form of particle-particle interaction. Users of FDPS can
write their programs by specifying the data structure of particles
they use, and calling necessary FDPS functions for domain
decomposition, particle migration between processes, and interaction
calculation. Currently, users should provide optimized function for
particle-particle interaction calculation.

Many of the parallel algorithms we newly implemented are not specific
to planetary rings but can be applied to any other particle-based
simulations. Using FDPS, users can write their programs in their
favorite language (currently C++, Fortran and C are supported)
\citep{2018PASJ...70...70N}, and let FDPS do complex parallelization.

Thus, it seems that one way to achieve performance and program
portability on new machines with rather exotic architecture such as
the machines evaluated in this paper is to develop the framework with
a common API and internal implementations specialized and optimized to
specific architectures. This is fairly straightforward in the case of
TaihuLight, in which the CPE and MPEs share the single physical
memory, since the data structure that FDPS handles can still in the
shared main memory. The basic data structure of FDPS is just an array of
particles, and both the user-developed  application program and the FDPS
side can access that particle array in usual way. 

On the other hand, how the separate memory spaces of PEZY-SC2 should
be handled within FDPS requires a bit more consideration. One
possibility would be to add the interface in which the user-side
programs, for example the function to perform time integration, is
passed to FDPS, instead of directly called within the user-side
program. Here, the function to be passed applies to single particle
(or some small array or particles), and applying it to all particles
in the system will be the responsibility of FDPS.  This approach will
probably make the software development and performance improvement
easier on other machines, since parallelization in both MPI and OpenMP
level can be taken care of within FDPS. 

This problem of portability is of course not limited to FDPS. It
occurs in practically any application in any field of
computational science. We clearly need a new and systematic approach
to solve this problem, and we think the use of frameworks such as
FDPS may be an efficient and practical way.

Our view of frameworks is that they should allow users to express
their problems in simple and (ideally) machine-independent way. In the
case of particle-based simulations, we have designed FDPS to meet this
goal, and it actually works pretty well on large HPC systems, both
with and without GPGPUs. Our experience on TaihuLight and PEZY-SC2
indicates that it is also possible to extend FDPS to cover these
systems. We believe similar approaches will be used in other fields.

\subsection{Summary}

In this paper, we described the implementation and performance of
a highly efficient simulation code for self-gravitating planetary
rings on  Sunway TaihuLight and PEZY-SC2 based systems.

The measured performance is 47.9 PF, or 39.7\% of the theoretical
peak, for simulation of $1.6\times 10^{12}$ particles on 40,000 nodes
of TaihuLight, 10.6PF, or 23.3\% of the theoretical peak, for
simulation of $8\times 10^{10}$ particles on 8192 nodes of Gyoukou,
and 1.01PF, or 35.5\% of the theoretical peak for $5\times 10^{9}$
particles on 512 nodes of Shoubu System B. As noted earlier, Gyoukou
and Shoubu System B use the same PEZY-SC2 processor. The difference in
the efficiency is purely due to the fact that Gyoukou was turned off
on March 31, 2018. The software at that time was still under
development.

Compared to previous achievements on K computer or ORNL Titan, the
achieved efficiency is similar or higher, and the speed in terms of
the number of particles integrated per second is higher, for both
TaihuLight and PEZY-SC2 based systems. As we stated earlier, this
level of performance would not be achieved without the new algorithms
described in this paper.

Compared to other multi-core processors for modern  HPC systems such
as Fujitsu SPARC64 VIIIfx and IXfx or Intel Xeon Phi processors,
both SW26010 processor  of TaihuLight and PEZY-SC2 processor of
PEZY-SC2 based systems have several unique features which allow very high peak
performance but at the same time make it much harder to achieve high
efficiency on real applications. These are:

\begin{itemize}

  \item Heterogeneous architecture with rather extreme performance
    ratio of 1:64 in the case of SW26010 and even larger in the case
    of SC2.
  \item The lack of cache hierarchy (SW26010) or cache coherency (SC2).
  \item Very limited main memory bandwidth, with B/F values around
    0.02--0.04.  This is about 1/10 of the numbers of Fujitsu or Intel HPC processors.
    
\end{itemize}  

On the other hand, SW26010 comes with very well-thought features which
allows the programmers to optimize the performance of code on
CPE. These features include:
\begin{itemize}

\item Low-latency DMA controller which can be initiated by any CPE.
\item Low-latency, high-bandwidth communication between CPEs.
  
\end{itemize}  

These two features allow very efficient use of the main memory
bandwidth. The two-dimensional structure of the network within CG seem
to be optimized for highly efficient implementation of matrix-matrix
multiplications, but it is actually quite useful for other real
applications, whenever fast inter-core communication is necessary.

It is certainly true that the need to use DMAs for data transfer
between CPE and main memory complicates the use of CPE. However, it is
also true that it makes quite optimized access to main memory
possible, since the application programmer can (or have to) control
all main memory accesses. In the case of our code, in several places
we have ``vectorizable'' loops, which perform the same operation on all
particles in the system. The number of operations per particle is
relatively small, of the order of ten, and the data size of one
particle is 32 bytes. In the case of manycore architecture with
hierarchical cache memory, to achieve high efficiency on simple vector
operations like 

{\tt a[i] =   b[i]+c[i]}

is actually quite complicated. In modern processors, load address
would be predicted and hardware prefetch is generated. The hardware
prefetch would probably work for a very simple loop like the above
example, but would fail if many vectors are loaded. Then the
programmer need to experiment with software prefetch, to find the way
to get  the best performance. 

In the case of SW26010, currently it is rather tedious and error-prone
to write the equivalent operation using the combination of Athread and
DMA, and sometimes inner kernel in assembly language, but once we do
so, we can get a performance close to the theoretical limit
relatively easily.

The existence of low-latency (less than 10 clock cycles) communication
path between CPEs is quite important for using CPEs for fine-grain
parallelism such as loop-level parallelization. Such low-latency
communication is difficult to implement on shared memory processors
with hierarchical cache.

The SC2 processor supports the cache flush and synchronization at each
level of the cache hierarchy, making the relatively low-latency
communications between processors possible.  However, it is clearly
desirable to have more direct control of interprocessor communication.

One common problem of SW26010 or SC2 is that writing high-performance
kernel for them means writing the inner kernel in the assembly
language. This is purely the software limitation, and probably not so
difficult to fix. In this aspect, SC2 is somewhat easier to deal with,
since it supports 8-way multithreaded execution, which effectively
hides the latencies of L1 cache and arithmetic unit from compiler.

In conclusion, we have implemented parallel particle simulation code
on Sunway SW26010 and PEZY-SC2 processors, and found that it is not
impossible to achieve high performance on their rather extreme
architectures with carefully designed algorithms. Even though the B/F
number are less than 0.1 and the network bandwidth is similarly low,
the efficiency we have achieved is comparable to that on K computer,
with 15 times more memory and network bandwidth. We feel that
architecture evolution in this direction will help the HPC community
to continue improving the performance. We also believe that high-level
software framework such as our FDPS will help many researchers to run
their own applications efficiently on new architectures.


\section*{Acknowledgment}

This work was supported by The Large Scale Computational Sciences with
Heterogeneous Many-Core Computers in grant-in-aid for High Performance
Computing with General Purpose Computers in MEXT (Ministry of
Education, Culture, Sports, Science and Technology-Japan), by JSPS
KAKENHI Grant Number JP18K11334 and JP18H04596 and also by JAMSTEC,
RIKEN, and PEZY Computing, K.K./ExaScaler Inc. In this research
computational resources of the PEZY-SC2 based systems supercomputer,
developed under the Large-scale energy-efficient supercomputer with
inductive coupling DRAM interface project of NexTEP program of Japan
Science and Technology Agency, has been used.


\bibliographystyle{SageH}
\bibliography{allrefs}

\end{document}